\documentclass[%
 reprint,
 amsmath,amssymb,
 aps,
]{revtex4-1}

\usepackage{graphicx}
\usepackage{dcolumn}
\usepackage{bm}

\begin{document}

\preprint{APS/123-QED}

\title{Unbound motion of massive particles in the Schwarzschild metric: \\
Analytical description in case of strong deflection}

\author{Oleg Yu. Tsupko}
 \email{tsupko@iki.rssi.ru}
\affiliation{%
 Space Research Institute of Russian Academy of Sciences, Profsoyuznaya 84/32, Moscow 117997, Russia\\
 National Research Nuclear University MEPhI, Kashirskoe Shosse 31, Moscow 115409, Russia\\
}%

\date{\today}

\begin{abstract}
Deflection angles of massive test particles moving along an unbound trajectory in the Schwarzschild metric are considered for the case of large deflection. We analytically consider the strong deflection limit, which is opposite to the commonly applied small deflection approximation and corresponds to the situation when a massive particle moves from infinity, makes several revolutions around a central object and goes to infinity. For this purpose we rewrite an integral expression for the deflection angle as an explicit function of the parameters determining the trajectory and expand it. Remarkably, in the limiting case of strong deflection, we succeed in deriving for the first time the analytical formulas for deflection angles as explicit functions of parameters at infinity. In particular, we show that in this case the deflection angle can be calculated as an explicit function of the impact parameter and velocity at infinity beyond the usual assumption of small deflection.
\begin{description}
\item[PACS numbers] 04.20.-q - 04.20.Cv - 04.25.Nx - 04.20.Jb - 95.10.Ce - 95.10.Eg


\end{description}
\pacs{04.20.-q - 95.30.Sf - 98.62.Sb - 94.20.ws}
\end{abstract}

\pacs{04.20.-q - 95.30.Sf - 98.62.Sb - 94.20.ws}
\maketitle


\section{Introduction}

The motion of massive test particles in the Schwarzschild metric has been extensively studied throughout the years \cite{Hagihara1931, Darwin1959, Darwin1961, Bogorodsky1962, M-Pleb1962, Metzner1963, Zeld-Novikov, MTW, Chandra, LL2, Weinberg, MiroRodriguez1, MiroRodriguez2}. Depending on initial parameters, there are different types of trajectories possible.

Let us consider an unbound orbit with a massive particle moving from infinity to a central object and then again to infinity. If the impact parameter of the particle is large, the trajectory is an almost straight line with a deflection by the small angle $\hat{\alpha} \ll 1$ (see, for example, Ref. \cite{MTW}):
\begin{equation} \label{alpha-small}
\hat{\alpha} = \frac{R_S}{b} \left( 1 + \frac{1}{v^2} \right) , \quad R_S = 2M, \quad  G=c=1 \, .
\end{equation}
Here $R_S$ is the Schwarzschild radius, and $b$ and $v$ are the impact parameter and velocity of incident test particles, respectively.

Beyond this simple case, the deflection angle of a massive particle moving from infinity to a central object and then to infinity cannot be calculated in such an easy way. The exact formula for the deflection angle is written as an integral from $R$ (the distance of the closest approach) to infinity and can be expressed via elliptic integrals.

In this paper we pay our attention to another limiting case in which the deflection angle can be calculated analytically, the strong deflection limit ($\hat{\alpha} \gg 1$), which corresponds to the situation in which a massive particle moves from infinity, makes several revolutions around a central object and goes to infinity. The deflection angle goes to infinity when the distance of the closest approach tends to the radius of unstable circular orbit of the particle (which depends on initial parameters). Thus, a strong deflection limit implies that the parameters of the orbit are close to their values at this radius.

Analytical formulas for strong deflection limit were derived only for the case of photon motion \cite{Darwin1959}. The deflection angle diverges logarithmically when $R$ approaches $3M$. For photon deflection beyond the small deflection limit, see classical papers and books \cite{Bogorodsky1962}, \cite{M-Pleb1962}, \cite{Metzner1963}, \cite{Zeld-Novikov}, \cite{MTW}, \cite{Chandra}, \cite{LL2}. The subject of photon deflection in a strong deflection situation is very popular in gravitational lensing theory. Photons that perform several revolutions around the central object can form so-called relativistic images of source. For gravitational lensing in the strong deflection case, see Refs. \cite{Virbhadra2000}, \cite{Bozza2001}, \cite{Perlick2004a}, \cite{BKTs2008}, \cite{Perlick2010}, \cite{Virbhadra2001}, \cite{Virbhadra2009}, \cite{PerlickMG}, \cite{Bozza2002}, \cite{Bozza2010}, and \cite{TsBK2013}.

The integral in the exact formula for the deflection angle of massive particles includes a polynomial of the third order in the radicand, and the expression via elliptic integrals is usually being written using the roots of this polynomial. In the general case, these three roots are different and supposed to be found numerically for given external parameters $E$ (energy at infinity per unit rest mass) and $L$ (angular momentum per unit rest mass), which determine the trajectory and $R$. For given $E$ and $L$, one has to obtain the roots of the polynomial numerically before being able to calculate the deflection angle (see, for example, Refs. \cite{M-Pleb1962} and \cite{Metzner1963}). Thus, if the impact parameter is not large, the deflection angle is not expressed explicitly via external parameters $E$ and $L$.

The divergence of the deflection angle leads to difficulties in the derivation of asymptotic formulas for large angles, and it seems that the most appropriate way is to expand the exact integral expression. For this purpose we need the integral to be an explicit function of parameters determining the trajectory. To achieve it, we suggest using either the pair ($R$, $L$) or ($R$, $E$) as an independent pair of parameters, instead of ($E$, $L$). In this case, all roots can be found analytically, and therefore we can perform all the calculations analytically. In this paper we obtain the deflection angle in the form of elliptic integrals as an explicit function of the two parameters ($R$, $L$) or ($R$, $E$). A similar way is used in the consideration of photon deflection, in which a similar polynomial is expressed via $R$.

Expanding the integrals in the asymptotic case of large deflection angles, we obtain the formulas for the deflection angles as analytic functions of ($R$, $L$) or ($R$, $E$). Nevertheless, for applications it is more convenient to determine the trajectory by the parameters at infinity. The distance of the closest approach $R$ is not a parameter at infinity, and therefore it is not good as an external parameter. Luckily, in the strong deflection limit, it becomes possible to express the deflection angle in terms of the impact parameter $b$ instead of $R$. Thus, only the parameters at infinity are used for calculation of the deflection angle, which is convenient for applications. A similar approach was used in the description of photon deflection in Ref. \cite{Bozza2001}.

In this work we present for the first time the analytical formulas for the deflection angles of massive particles in the strong deflection limit ($\hat{\alpha} \gg 1$) as an explicit function of external parameters at infinity, ($b$, $L$) or ($b$, $E$). Parameters at infinity are simply related with each other, so it is possible to calculate the deflection angle with our formulas in very different situations. For example, our formulas allow one to calculate analytically the deflection angle of a particle with a given impact parameter and velocity at infinity as initial conditions, beyond the usual small deflection case.

Results of the present paper can be useful for the investigation of star motion around supermassive black holes. In this case, the ratio of masses is extremely high, and such orbits now are often referred to as extreme mass ratio inspirals (EMRIs). Interest in these orbits is very high now, since they are some of the most prominent sources of gravitational waves (see, for example, the pioneering work of Zel'dovich and Novikov \cite{zeld-nov1964}). With appropriate initial conditions, the test body can perform several revolutions around the central object. In this case, weak field approximation is not valid, and the usual way is to solve the strong field equations numerically. Our formulas allow an easy explicit calculation of the angular deflection of a star or compact object in EMRIs and then finding, for example, the number of revolutions of the test body around a supermassive black hole for various initial conditions, which can be useful for gravitational wave estimation \cite{zeld-nov1964}.

In our recent work \cite{TsBK2013}, we considered unbound photon orbits in the Schwarzschild metric in presence of homogenously distributed plasma, in the case of strong deflection. In particular, we have derived the formulas for the deflection angles, in the case of strong deflection. In the paper of Kulsrud and Loeb \cite{Kulsrud-Loeb}, it was shown that in homogeneous plasma the photon wave packet moves like a massive particle with velocity equal to the group velocity of the wave packet, the rest mass equal to the plasma frequency, and its energy equal to the photon energy. This analogy allows one to apply the results of that paper to the calculation of deflection angle of massive particle, with a corresponding change of variables. In some places of the present paper (where possible), we will refer to some results of that paper. For the subject of gravitational lensing in plasma, see also Refs. \cite{BliokhMinakov}, \cite{Perlick2000}, \cite{BKTs2009}, \cite{BKTs2010}, \cite{morozova}, \cite{mao}, and \cite{Bicak}.

The paper is organized as follows. In Sec. II we derive the well-known exact expression for the deflection angle of the massive test particle moving in the Schwarzschild metric along an unbound orbit. In Sec.III we express the formula for deflection via elliptic integrals, working first with the pair of independent parameters $(R, L)$ and then, separately, with the pair $(R, E)$. In Sec. IV we find a critical value for the closest approach distance at which a massive particle going from infinity remains on the circular orbit with an infinite number of circles around the center. This critical value depends on $L$ (or on $E$). The deflection angle goes to infinity when the distance of the closest approach goes to this value. In Sec. V we derive an analytical formula for the particle deflection angle in the Schwarzschild metric, in the limit of strong deflection. In Sec. VI we conclude and discuss our results.

\section{Exact deflection angle}

Let us consider a massive test particle moving in the Schwarzschild metric:
\begin{equation}
ds^2 =  - A(r) \, dt^2 + \frac{dr^2}{A(r)} + r^2
\left( d \theta^2 + \sin^2 \theta d \varphi^2 \right) ,
\label{metric}
\end{equation}
$$
A(r) = 1 - \frac{2M}{r}.
$$
The system of units used here is
\begin{equation}
G=c=1, \quad \mbox{the Schwarzschild radius} \;\; R_S=2M.
\end{equation}
Indices are
\begin{equation}
i,k = 0,1,2,3; \quad \alpha, \beta = 1,2,3 \; (r, \theta, \varphi) \, ,
\end{equation}
\begin{equation}
\mbox{ signature } \; \; \{-,+,+,+\} .
\end{equation}

Equations governing the orbit of the test massive particle (see, for example, Ref. \cite{MTW}) are:
\begin{equation}
\left( \frac{dr}{d\tau} \right)^2 = E^2 - V^2(r) \quad \mbox{(radial part of motion)},
\end{equation}

\begin{equation}
\frac{d \varphi}{d\tau} = \frac{L}{r^2} \quad \mbox{(angular part of motion)}.
\end{equation}
Here $E$ is the energy at infinity per unit rest mass of the particle, $L$ is the angular momentum per unit rest mass of the particle, and the effective potential $V(r)$ is
\begin{equation}
V^2(r) = \left( 1-\frac{2M}{r}  \right) \left( 1+ \frac{L^2}{r^2}   \right) .
\end{equation}

From equations of orbit, we can obtain the equation of the trajectory as
\begin{equation} \label{trajectory}
\frac{d \varphi}{dr} = \pm \frac{L}{r^2} \frac{1}{\sqrt{E^2 - A(r) \left( 1 + \frac{L^2}{r^2} \right) }}.
\end{equation}

Assume that the test particle moves in such a way that its $\varphi$ coordinate increases. Then the plus sign in Eq. (\ref{trajectory}) corresponds to motion with the coordinate $r$ also increasing, and the minus sign corresponds to motion with decreasing $r$.

For a particle which moves from infinity to the distance of the closest approach $R$ (minimal value of the coordinate $r$) and then to infinity, a change of angular coordinate is

\begin{equation}
\Delta \varphi = - \int \limits_\infty^R \frac{L}{r^2} \, \frac{dr}{\sqrt{E^2 - A(r) \left( 1 + \frac{L^2}{r^2} \right)} }    \; +
\end{equation}
$$
+ \int \limits_R^\infty \frac{L}{r^2} \, \frac{dr}{\sqrt{E^2 - A(r) \left( 1 + \frac{L^2}{r^2} \right)} } =
$$
$$
=
2 \int \limits_R^\infty \frac{L}{r^2} \, \frac{dr}{\sqrt{E^2 - A(r) \left( 1 + \frac{L^2}{r^2} \right)} } \, .
$$
The motion along the straight line corresponds to the change of the angular coordinate $\Delta \varphi = \pi$.
Therefore the deflection angle can be written as
\begin{equation} \label{angle-EL}
\hat{\alpha} =  2 \int \limits_R^\infty \frac{L}{r^2} \frac{dr}{\sqrt{E^2 - A(r) \left( 1 + \frac{L^2}{r^2}  \right) }} - \pi .
\end{equation}

The closest approach distance $R$ and the parameters $E$ and $L$ are connected at the point $r=R$ by the following boundary condition:
\begin{equation} \label{boundary-cond}
E^2 = A(R) \left( 1 + \frac{L^2}{R^2} \right) .
\end{equation}
The trajectory and the deflection angle are thus completely determined by any two parameters from \{$R$,  $E$, $L$\}, with the
third parameter being expressed through Eq. (\ref{boundary-cond}). To compare, the photon motion is determined by only one parameter, either $R$ or $b$, which are uniquely connected with each other.

\section{Expression of the deflection angle via elliptic integrals}

Here we express the integral for the angle (\ref{angle-EL}) via elliptic integrals.

Let us introduce notations
\begin{equation}
\frac{1}{r} = u, \; \frac{1}{R} = u_0, \; A(u) = 1 - 2Mu, \; A(u_0) = 1 - 2Mu_0,
\end{equation}
and rewrite Eq. (\ref{angle-EL}) in terms of $u$ instead of $r$. We will refer to the expression in radicand in Eq. (\ref{angle-EL}) as
$f(u)$. Equation (\ref{angle-EL}) becomes
\begin{equation}
\label{alpha-with-f}
\hat{\alpha} =  2 \int \limits_0^{u_0} \frac{L du}{\sqrt {f(u)}} - \pi \, ,
\end{equation}
\begin{equation} \label{fu}
\mbox{where} \quad f(u) = E^2 - (1-2Mu)(1+L^2 u^2) \, .
\end{equation}
Coefficients of the polynomial $f(u)$ are functions of $E$ and $L$, and the roots of this polynomial could be found either numerically or by a very complicated analytical solution of the third-order algebraic equation. We suggest using $R$ and $L$ (or $R$ and $E$) as another pair of independent parameters. Using the boundary condition (\ref{boundary-cond}), we write
\begin{equation}
E^2 = (1-2Mu_0)(1+L^2 u_0^2) \, ,
\end{equation}
\begin{equation}
f(u) = (1-2M u_0)(1+ L^2 u_0^2) - (1-2Mu)(1+L^2 u^2) \, .
\end{equation}
Now the coefficients of the polynomial are functions of $R$ and $L$, and one of the roots $u=1/R$ is evidently visible. The other two roots (as functions of $R$ and $L$) can be found easily by solving the quadratic equation. This approach is productive because we can perform all further calculations analytically and obtain the deflection angle in the form of elliptic integrals as an explicit function of the pair of parameters $R$ and $L$. A similar approach is usually applied for the deflection of photons in vacuum, where an analogous polynomial is expressed via $R$ (see, for example, Ref. \cite{BKTs2008} and other references in the present paper). We would like to mention that in the case that is of relevance for the subject of the paper the polynomial $f(u)$ has three real roots, and $R$ is the biggest of the three values for which $f(1/R)=0$. Finally,
\begin{equation}
f(u) = 2ML^2 (u-u_A)(u-u_B)(u-u_C) ,
\end{equation}
where
\begin{equation} \label{u-ABC}
u_A = \frac{R-2M + Q  }{4MR} \, , \; u_B = u_0 = \frac{1}{R} \, , \; u_C = \frac{R-2M - Q  }{4MR} \, ,
\end{equation}
\begin{equation} \label{definition-Q-RL}
Q^2 = (R-2M)^2 + 8M (R-2M) \, \left( 1 - \frac{2MR^2}{L^2(R-2M)} \right) \, .
\end{equation}
For the deflection angle $\hat{\alpha}$, we obtain the expression
\begin{equation}
\hat{\alpha} = \frac{2}{\sqrt{2M}} \int \limits_0^{u_0} \frac{du}{\sqrt{(u-u_A)(u-u_B)(u-u_C)}}  -  \pi \, .
\label{angle-RL-korni}
\end{equation}
Using Ref. \cite{Gr-Ryzhik}, we write
\begin{equation} \label{Gr-R}
\int \limits_0^{u_B} \frac{du}{\sqrt{(u-u_A)(u-u_B)(u-u_C)}} = \frac{2}{\sqrt{u_A-u_C}} \, F(z, k) \, ,
\end{equation}
$$
z = \sqrt{\frac{(u_A-u_C)u_B}{(u_B-u_C)u_A}} \, , \quad k = \sqrt{\frac{u_B-u_C}{u_A-u_C}} \, ,
$$
$$
[u_A > u_B > 0 \geq u_C] \, .
$$
Here $F(z,k)$ is an elliptic integral of the first kind \cite{Korn, Gr-Ryzhik}:
\begin{equation}
F(z,k) = \int \limits_0^z \frac{dx}{\sqrt{(1-x^2)(1-k^2 x^2)} } =  \int \limits_0^\varphi \frac{d \theta}{\sqrt{1 - k^2 \sin^2 \theta}} =
\end{equation}
$$
=  \tilde{F} (\varphi, k) \, ,  \; F(\sin \varphi, k) \equiv \tilde{F} (\varphi, k) \, ,  \;  x = \sin \theta \, , \;  z=\sin \varphi \, .
$$
Using Eq. (\ref{Gr-R}) in Eq. (\ref{angle-RL-korni}) and substituting Eq. (\ref{u-ABC}), we obtain the deflection angle $\hat{\alpha}$ in the form
\begin{equation} \label{alpha-exact1}
\hat{\alpha} = 4 \sqrt{\frac{R}{Q}} \, F(\tilde{y}, k)  -  \pi  \, ,
\end{equation}
\begin{equation}
\mbox{where} \quad \tilde{y} = \sqrt{  \frac{8MQ}{( 6M-R+Q )( R-2M+Q )} }  \, ,
\label{y}
\end{equation}
\begin{equation} \label{k}
k = \sqrt{ \frac{6M-R+Q}{2Q} } \, .
\end{equation}
The expression (\ref{alpha-exact1}) can be also written differently, due to the property of elliptic integrals \cite{Abramowitz}, which in our notations reads
\begin{equation}
F(z,k) + F(\tilde{y},k) = F(1,k) \, ,
\label{zy}
\end{equation}
$$
\mbox{provided} \quad \sqrt{1-k^2} \, \frac{z}{\sqrt{1-z^2}} \, \frac{\tilde{y}}{\sqrt{1-\tilde{y}^2}} = 1 \, .
$$
Here $F(1,k) = \tilde{F}(\pi/2,k) = K(k)$ is the complete elliptic integral of the first kind. It is easy to check that
\begin{equation}
\label{z}
z^2 = \frac{2M+Q-R}{6M+Q-R} \,
\end{equation}
and $\tilde{y}$ given by Eq. (\ref{y}) satisfy the relation (\ref{zy}), independently on the form of $Q$. Therefore, we obtain the deflection angle $\hat{\alpha}$ in the form

\begin{equation} \label{alpha-exact2}
\hat{\alpha} = 4 \sqrt{\frac{R}{Q}} \, \left[ F(1,k) - F(z,k) \right]   -  \pi.
\end{equation}
Thus, we have obtained, for the first time, the formulas for the deflection angle (\ref{alpha-exact1}), (\ref{alpha-exact2}) of massive test particles, where arguments of elliptical integrals are expressed explicitly via parameters $R$ and $L$, defining the trajectory.

To obtain the deflection angle as a function of $R$ and $E$, we rewrite Eq. (\ref{boundary-cond}) as
\begin{equation} \label{bond-cond-L}
L^2 = R^2 \left( \frac{E^2}{A(R)} - 1 \right) \, ,
\end{equation}
and substitute Eq. (\ref{bond-cond-L}) into the expression for the deflection angle (\ref{angle-EL}). The deflection angle then will be in the form of Eq. (\ref{alpha-with-f}) with the expression $f(u)$ in the form
\begin{equation}
f(u) = \frac{2M R^2  \, (R E^2 - R+2M)}{R-2M}  \, (u-u_A) (u-u_B) (u-u_C) \, ,
\end{equation}
where
\begin{equation} \label{u-ABC-RE}
u_A = \frac{R-2M + Q  }{4MR} \, , \; u_B = u_0 = \frac{1}{R} \, , \; u_C = \frac{R-2M - Q  }{4MR} \, ,
\end{equation}
\begin{equation} \label{definition-Q-RE}
Q^2 = (R-2M)^2 + 8M (R-2M) \, \frac{1}{1+ \frac{2M}{R(E^2-1)}} \, .
\end{equation}
Finally, in this case we obtain the same expressions (\ref{alpha-exact1}) and (\ref{alpha-exact2}) with the only difference in the definition of $Q$. For more details about this case, see our work in Ref. \cite{TsBK2013}.

We obtain the exact expression for the deflection angle as an explicit function of the pair $(R, L)$ or $(R, E)$, see Eqs. (\ref{alpha-exact1}) and (\ref{alpha-exact2}), with the variables $\tilde{y}$ given by Eq. (\ref{y}), $k$ by Eq. (\ref{k}), $z$ by Eq. (\ref{z}), and $Q$ in the form of Eq. (\ref{definition-Q-RL}) for the $(R, L)$ case or Eq. (\ref{definition-Q-RE}) for the $(R, E)$ case.

The exact expression allows one to calculate the deflection angle for any ranges of $\hat{\alpha}$. In case of small deflection angles ($\hat{\alpha} \ll 1$), there is an analytical formula for the deflection angle (\ref{alpha-small}). In the opposite case of strong deflection ($\hat{\alpha} \gg 1$), we can also obtain an asymptotic analytical formula for the deflection angle, expanding formula (\ref{alpha-exact2}) for the exact deflection.

\section{Critical distance of the closest approach for unbound massive test particles}

To obtain an expression for the angle in the limit of strong deflection, we need to expand the integral (\ref{alpha-exact2}) about the value of $R$, at which the deflection angle goes to infinity. It will be referred to as the critical value of $R$.

The deflection angle (\ref{alpha-with-f}) can be written in terms of an effective potential \cite{MTW},
\begin{equation}
\hat{\alpha} = 2 \int \limits_0^{u_0} \frac{L du}{\sqrt{E^2 - V^2(u)}} - \pi \, ,
\end{equation}
where
\begin{equation}
V^2(u) = (1-2Mu) ( 1 + L^2 u^2 ) \, .
\end{equation}
From the condition
\begin{equation}
\frac{dV^2(u)}{du} = 0
\end{equation}
we obtain
\begin{equation}
u = \frac{1 \pm \sqrt{1-12M^2/L^2}}{6M}  \, .
\end{equation}
The maximum and minimum of the effective potential are located at the points $r_M$ and $r_m$, respectively:
\begin{equation} \label{rM-L}
r_M = \frac{6M}{1 + x} \, ,  \;   r_m = \frac{6M}{1 - x} \, , \; x \equiv \sqrt{1-\frac{12M^2}{L^2}}, \; r_M < r_m \, .
\end{equation}

The point $r=r_M$ corresponds to an unstable circular orbit of the particle, and $r_m$ corresponds to a stable circular orbit. The critical distance of the closest approach (at given $L$) is being reached when the value of $R$ coincides with $r_M$. In other words, the unbound orbit with infinite deflection takes place if $R=r_M$. For simplicity, we will denote the critical value of the closest approach as $r_M$.

The critical distance of the closest approach $r_M$ is a function of $L$. To derive the strong deflection limit as a function of $(R, L)$, one should expand the integrals in exact formulas for deflection angles near $R=r_M$ where $r_M$ is expressed in terms of $L$. Analogously, to get the strong deflection limit as a function of $(R, E)$,one needs to rewrite $r_M$ as a function of $E$ instead of $L$.

Using Eqs. (\ref{bond-cond-L}) and (\ref{rM-L}), we express the critical distance $r_M$ for unbound massive particles coming from infinity as a function of $E$ in the form
\begin{equation} \label{rM-E-with-y}
r_M = \frac{3E^2 - 4 + 3E^2 y}{2(E^2-1)} \, M \, , \quad  y \equiv \sqrt{1-\frac{8}{9E^2}} \, .
\end{equation}
Noticing that $3E^2-4+3E^2 y =\frac{3}{2} E^2 (3y-1)(1+y)$, we write $r_M$ as
\begin{equation}
r_M = \frac{3}{4} \, \frac{(3y-1)(1+y)}{1-\frac{1}{E^2}} \, M \, .
\end{equation}
Then, because of the equality $E^2(3y-1)(3y+1)=8(E^2-1)$, $r_M$ takes the following form:
\begin{equation} \label{rM-E}
r_M = 6M \, \frac{1+y}{1+3y} \, , \quad  y \equiv \sqrt{1-\frac{8}{9E^2}} \, .
\end{equation}

At given $L$ (or $E$), the trajectory is determined by the choice of the second parameter $R$. The deflection angle goes to infinity when $R$ approaches $r_M$, and the massive particle performs an infinite number of turns at the radius $r_{M}$. Therefore, $r_{M}$ is the critical (minimal) distance of the closest approach $R$, for given $L$ (or $E$). For unbound orbits of massive particles coming from infinity, making several loops around a black hole and going back to infinity, the value of the critical distance of the closest approach is between $3M$ and $4M$.

We should mention here that the points of maximum $r_M$ and minimum $r_m$ of the effective potential appear only for $L> 2\sqrt{3}M$. According to Eqs. (\ref{rM-L}) and (\ref{rM-E}), $r_M$ varies from $6M$ to $3M$ if $L$ varies from $2 \sqrt{3} M$ to infinity or, equivalently, if $E$ varies from
$\sqrt{8}/3$ to infinity. In this paper we consider only the case in which the particle moves from infinity; therefore, for the studied orbit, it has $E \geq 1$. As we are interested in the case in which the particle reaches the point $R$ and then returns to infinity, we need $V(r_M)$ to be more than unity. In another words, the point of the maximum of the effective potential should be higher than the value of effective potential at infinity. This happens if $L \geq 4M$ (see Ref. \cite{MTW}). Therefore, for the studied orbit, $r_M$ varies from $4M$ to $3M$ if $L$ varies from $4 M$ to infinity or, equivalently, if $E$ varies from $1$ to infinity.

\section{Deflection angle in the strong deflection limit}

\subsection{As a function of $R$ and $L$}

When $R$ tends to $r_M$ ($R>r_M$), the deflection angle goes to infinity. To obtain the analytic expression for the deflection angle in the strong deflection limit ($\hat{\alpha} \gg 1$), one needs to expand the integral (\ref{alpha-exact2}) around the point $R=r_M$.

As follows from Eqs. (\ref{k}) and (\ref{definition-Q-RL}), $k \simeq 1$ corresponds to $R \simeq r_M$, so we can use the expansions of the elliptic integrals about $k=1$ \cite{Zhuravsky1941, Gr-Ryzhik}:
\begin{equation} \label{eq42}
F(1,k) \simeq \ln \frac{4}{\sqrt{1-k^2}} \, ,
\end{equation}
\begin{equation} \label{eq43}
F(z,k) \simeq \ln \tan \left( \frac{\arcsin(z)}{2} + \frac{\pi}{4} \right) \, .
\end{equation}
The expression (\ref{alpha-exact2}) for the deflection angle is simplified to the following:
\begin{equation} \label{alpha-prelim}
\hat{\alpha} = 4 \sqrt{\frac{R}{Q}} \left[ \ln \frac{4}{\sqrt{1-k^2}} - \ln \frac{1+z}{\sqrt{1-z^2}}  \right] - \pi \, .
\end{equation}
Expanding ($1-k^2$) near $R=r_M$ [$r_M$ is defined by Eq. (\ref{rM-L})], after simplifications (see Appendix A) we have
\begin{equation} \label{1-k2-RL}
1-k^2 \simeq \frac{(1+x)^2}{9Mx} (R-r_M) \, .
\end{equation}
Thus, expanding Eq. (\ref{alpha-prelim}) near $R=r_M$ and keeping the leading terms, we obtain
\begin{equation} \label{alpha-with-z}
\hat{\alpha} = - \frac{2}{\sqrt{x}} \ln \left[\frac{(R-r_M)M}{4 r_M^2 x} \, \frac{(1+z)^2}{1-z^2}  \right] -
\pi \, ,
\end{equation}
where $z$ [expression (\ref{z}) with $Q$ taken in the form of Eq. (\ref{definition-Q-RL})] is simplified to
\begin{equation} \label{z-new-L}
z = \sqrt{\frac{2x-1}{3x}} \, .
\end{equation}
Substituting $z$, we obtain the deflection angle $\hat{\alpha}$ as
a function of the closest approach distance $R$ and $L$ in the form
\begin{equation} \label{alpha-R-L}
\hat{\alpha}(R,L) = -\frac{2}{\sqrt{x}}  \ln \left[z_1(x) \, \frac{R-r_M}{r_M}  \right] - \pi \, ,
\end{equation}
where
\begin{equation} \label{z-x-RL}
\quad z_1(x) = \frac{5x-1+2\sqrt{3x(2x-1)}}{24x} \, ,
\end{equation}
and $r_M$ and $x$ are defined in Eq. (\ref{rM-L}).
This formula (\ref{alpha-R-L}) is asymptotic and valid for $R$ close to $r_M$.

\subsection{As a function of $R$ and $E$}

In this case, it follows from Eqs. (\ref{k}) and (\ref{definition-Q-RE}) that $k \simeq 1$ corresponds to $R \simeq r_M$, so we use the same formulas (\ref{eq42}), (\ref{eq43}), (\ref{alpha-prelim}). Expanding ($1-k^2$) near $R=r_M$ [$r_M$ is given by (\ref{rM-E})], after significant simplifications (see Ref. \cite{TsBK2013} for details) we have
\begin{equation} \label{1-k2-RE}
1-k^2 \simeq 2 \, \frac{1+y}{y} \, \frac{M(R-r_M)}{r_M^2} \, .
\end{equation}
Expanding Eq. (\ref{alpha-prelim}) near the point $R=r_M$ and keeping the leading terms, we obtain
\begin{equation} \label{alpha-with-z}
\hat{\alpha} = -2 \sqrt{\frac{1+y}{2y}} \ln \left[ \frac{1+y}{8y} \, \frac{(R-r_M)M}{r_M^2} \, \frac{(1+z)^2}{1-z^2}  \right] -
\pi \, ,
\end{equation}
where $z$ [expression (\ref{z}) with $Q$ taken in the form of Eq. (\ref{definition-Q-RE})] is simplified to
\begin{equation} \label{z-new-E}
z = \sqrt{\frac{3y-1}{6y}} \, .
\end{equation}
Substituting $z$, we obtain the deflection angle $\hat{\alpha}$ as a function of the closest approach distance $R$ and $E$ in the form
\begin{equation} \label{alpha-R-E}
\hat{\alpha}(R,E) = -2 \sqrt{\frac{1+y}{2y}} \ln \left[z_2(y) \, \frac{R-r_M}{r_M}  \right] - \pi \, ,
\end{equation}
where
\begin{equation} \label{z-x-RE}
\quad z_2(y) = \frac{9y-1+2\sqrt{6y(3y-1)}}{48y} \, ,
\end{equation}
and $r_M$ and $y$ are defined in Eq. (\ref{rM-E}).
This formula (\ref{alpha-R-E}) is asymptotic and valid for $R$ close to $r_M$.

\subsection{As a function of $b$ and $L$}

The impact parameter $b$ for a massive particle is defined uniquely by $E$ and $L$ as \cite{MTW}, \cite{Weinberg}
\begin{equation} \label{definition-b}
b^2 = \frac{L^2}{E^2-1} \, .
\end{equation}

Let us assume that $b_{cr}$ is the minimal impact parameter, corresponding to the critical distance of the closest approach $r_M$. Since we need $b$ to be a function of $R$ and $L$, we substitute $E^2$ from Eq. (\ref{boundary-cond}) into Eq. (\ref{definition-b}). Expanding $b$ in Eq. (\ref{definition-b}) around the point $R=r_M$, we obtain
\begin{equation} \label{b-L}
b = b_{cr}(x) + b_1 (R-r_M)^2 \, ,
\end{equation}
where
\begin{equation} \label{b-cr-L}
b_{cr}(x) = r_M \, \sqrt{\frac{3}{2x-1}} \, ,
\end{equation}
\begin{equation}
b_1 = \frac{3x}{2 (2x-1) r_M^2} \, b_{cr}(x) \, .
\end{equation}
Writing
\begin{equation}
\frac{(R-r_M)^2}{r_M^2} = 2z^2 \, \frac{b-b_{cr}(x)}{b_{cr}(x)} \, ,
\end{equation}
we obtain the deflection angle $\hat{\alpha}$ as a function of the impact parameter $b$ and $L$,
\begin{equation} \label{alpha-b-L}
\hat{\alpha}(b,L) = - \frac{1}{\sqrt{x}}  \, \ln \left[ 2 z^2 \, z_1^2(x) \, \frac{b-b_{cr}(x)}{b_{cr}(x)} \right] - \pi \, ,
\end{equation}
where $z$ is given by Eq. (\ref{z-new-L}), $z_1(x)$ is given by Eq. (\ref{z-x-RL}), $b_{cr}(x)$ is given by Eq. (\ref{b-cr-L}), and $r_M$ and $x$ are defined in Eq. (\ref{rM-L}). This formula is valid for $b$ close to $b_{cr}(x)$, where $b_{cr}(x)$ is the critical value of impact parameter at given $L$ (\ref{b-cr-L}).

\subsection{As a function of $b$ and $E$}

In this case we need $b$ to be a function of $R$ and $E$, and we substitute $L^2$ from Eq. (\ref{bond-cond-L}) into Eq. (\ref{definition-b}). Expanding $b$ in Eq. (\ref{definition-b}) around the point $R=r_M$, we obtain (for this case, see Ref. \cite{TsBK2013})
\begin{equation} \label{b-E}
b = b_{cr}(y) + b_1 (R-r_M)^2 \, ,
\end{equation}
where
\begin{equation} \label{b-cr-E}
b_{cr}(y) = \sqrt{3} \, r_M \, \sqrt{\frac{1+y}{3y-1}} \, ,
\end{equation}
\begin{equation}
b_1 = \frac{3\sqrt{3}}{2} \, \frac{y}{r_M} \, \sqrt{\frac{1+y}{3y-1}} = \frac{3y}{2r_M^2} \, b_{cr}(y) \, .
\end{equation}
Writing
\begin{equation}
\frac{(R-r_M)^2}{r_M^2} = \frac{2}{3y} \, \frac{b-b_{cr}(y)}{b_{cr}(y)} \, ,
\end{equation}
we obtain the deflection angle $\hat{\alpha}$ as a function of the impact parameter $b$ and $E$:
\begin{equation} \label{alpha-b-E}
\hat{\alpha}(b,E) = - \sqrt{\frac{1+y}{2y}} \, \ln \left[ \frac{2\,z_2^2(y)}{3y} \, \frac{b-b_{cr}(y)}{b_{cr}(y)} \right] - \pi \, ,
\end{equation}
where $z_2(y)$ is given by Eq. (\ref{z-x-RE}), $b_{cr}(y)$ is given by Eq. (\ref{b-cr-E}), and $r_M$ and $y$ are defined in Eq. (\ref{rM-E}). This formula is valid for $b$ close to $b_{cr}(y)$, where $b_{cr}(y)$ is the critical value of impact parameter at given $E$ (\ref{b-cr-E}).

If impact parameter $b$ and asymptotic velocity at infinity $v$ are used as initial parameters, the deflection angle can be calculated by the formula (\ref{alpha-b-E}) with
$$
E = \frac{1}{\sqrt{1-v^2}} \, .
$$
In other cases connection $L = Evb$ (see, for example, Ref. \cite{MTW}) can be used.

\subsection{Comparison of formulas}

We have shown the derivation of the formulas for the two cases \{($R$, $L$), ($b$, $L$)\} and \{($R$, $E$), ($b$, $E$)\} from the very outset to the final results. In the first case, we used $L$ as the second independent variable, and in the other case we used $E$ instead.

Here we would like to discuss also how these two cases can be transformed to each other by the appropriate change of variables.

In the case of formulas with the distance of the closest approach $R$, Eqs. (\ref{alpha-R-L}) and (\ref{alpha-R-E}), it is easy to do. Comparing Eqs. (\ref{rM-L}) and (\ref{rM-E}), one easily finds that $x=2y/(1+y)$. Substituting this into Eq. ($\ref{alpha-R-L}$), we obtain the deflection angle $\hat{\alpha}$ as a function of the closest approach distance $R$ and $E$ in the form (\ref{alpha-R-E}).

In the case of formulas with the impact parameter $b$, Eqs. (\ref{alpha-b-L}) and (\ref{alpha-b-E}), the change of variables is more complicated and cannot be performed by this simple substitution. In the derivation of these formulas, we expand the impact parameter up to quadratic terms $(R-r_M)^2$. Therefore, in this case it is necessary to perform the change of variables with the same accuracy. To get the expression (\ref{b-E}) from Eq. (\ref{b-L}), it is necessary to substitute the exact expression of $L$ via $R$ and $E$ [see Eq. (\ref{bond-cond-L})] to Eq. (\ref{b-L}) and expand it up to quadratic terms $(R-r_M)^2$. In this procedure, additional quadratic terms will arise from the term $b_{cr}(x)$ in Eq. (\ref{b-L}), and after simplifications we can finally get the expression (\ref{b-E}). It is also possible to get the expression (\ref{b-L}) from Eq. (\ref{b-E}). Therefore, we can transform formulas (\ref{alpha-b-L}) and (\ref{alpha-b-E}) into each other.

We should emphasize therefore that if one wants to use other variables instead of the pair used in the given formula one should calculate the new pair of variables using the exact relations, for example, Eq. (\ref{definition-b}).

In the limiting case of $L \rightarrow \infty$ or $E \rightarrow \infty$, our formulas for massive particles transform to formulas for photons in vacuum; for the deflection angle of photons in the strong deflection limit with $R$, see Ref. \cite{Darwin1959}, and for the formula with $b$, see Ref. \cite{Bozza2001}.

\section{Discussion}

The main result of present paper is a derivation of the analytical formulas (\ref{alpha-R-L}), (\ref{alpha-R-E}), (\ref{alpha-b-L}), (\ref{alpha-b-E}) for deflection angles of a massive test particle in the strong deflection limit ($\hat{\alpha} \gg 1$). The large deflection limit is the opposite of the commonly used small deflection case and works very well for the situation in which a massive particle goes from infinity, performs several revolutions around the central object ,and then goes to infinity. Formulas are written as functions of parameters at infinity determining the trajectory.

\section*{Acknowledgments}
OYuT has the honour of thanking G.S. Bisnovatyi-Kogan for motivation, discussion, and permanent support during the work on this paper.

We would like to thank anonymous referees for many useful remarks, especially concerning the ideas about the change of variables, which is discussed in Subsection E of Section V.

OYuT would like to thank N.S. Voronova for the help with brushing up the language of the present paper.

The work of OYuT was partially supported by the Russian Foundation for Basic Research Grant No. 12-02-31413 (project 'My First Grant'), the Russian Federation President Grant for Support of Young Scientists, Grant No. MK-2918.2013.2, and the Dynasty Foundation.

\appendix

\section{Calculations for Sec. V}
To find an expansion (\ref{1-k2-RL}) of ($1-k^2$) near the point $R=r_M$, let us rewrite $k^2$ as
\begin{equation} \label{zA-1}
k^2 = \frac{6M-R+Q}{2Q} = \frac{1}{2} + \frac{1}{2} \, \frac{6M-R}{Q} \, .
\end{equation}
For convenience, let us calculate $Q^2/(6M-R)^2$. Here $Q^2$ is
\begin{equation} \label{za-def-Q-L}
Q^2 = (R-2M)^2 + 8M (R-2M) \, \left( 1 - \frac{2MR^2}{L^2(R-2M)} \right) \, .
\end{equation}
Let us write $R$ as
\begin{equation}
R = r_M + \Delta R \, .
\end{equation}
Here $\Delta R = R - r_M$ is a small variable:
\begin{equation}
\Delta R \ll r_M \, ,
\end{equation}
and
\begin{equation} \label{zA-rM-x}
r_M = \frac{6M}{1 + x} \, .
\end{equation}

Expanding $Q^2$ near $R=r_M$ (small variable here is $\Delta R$), we obtain:
$$
Q^2 \simeq Q_0 + Q_1 \Delta R \, ,
$$
where
$$
Q_0 = \frac{36 M^2 x^2}{(1+x)^2}, \quad  Q_1 = \frac{4Mx(1+4x)}{1+x} \, .
$$
Expanding $(6M-R)^2$ near $R=r_M$ (small variable here is $\Delta R$), we obtain:
\begin{equation}
(6M-R)^2 \simeq \frac{36 M^2 x^2}{(1+x)^2} \left( 1 - \frac{1+x}{3Mx} \, \Delta R \right) \, .
\end{equation}
Keeping the leadings terms, we obtain $Q^2/(6M-R)^2$ in the form
\begin{equation}
\frac{Q^2}{(6M-R)^2} \simeq 1 + \frac{4(1+x)^2}{9Mx} \, \Delta R \, .
\end{equation}

With using (\ref{zA-1}) we obtain finally
\begin{equation}
1-k^2 \simeq \frac{(1+x)^2}{9Mx} \, \Delta R \, .
\end{equation}

%


\end{document}